\documentclass[aps,amsmath,amssymb,amsfonts,showpacs,showkeys]{revtex4}

\usepackage{epsfig,psfrag}

\newcommand{\beq}{\begin{equation}}
\newcommand{\eeq}{\end{equation}}
\newcommand{\bea}{\begin{eqnarray}}
\newcommand{\eea}{\end{eqnarray}}

\newcommand{\nn}{\nonumber\\}

\newcommand\eqn[1]{Eq.\,(\ref{#1})}

\newcommand\fig[1]{Fig.\,{\ref{#1}}}
\newcommand\sect[1]{Sect.\,{\ref{#1}}}

\newcommand\tab[1]{Table~\ref{#1}}

\begin{document}

\title{Optimized regulator for the quantized anharmonic oscillator}

\author{J. Kovacs}
\affiliation{Department of Theoretical Physics, University of Debrecen,
P.O. Box 5, H-4010 Debrecen, Hungary}

\author{S. Nagy}
\affiliation{Department of Theoretical Physics, University of Debrecen,
P.O. Box 5, H-4010 Debrecen, Hungary}
\affiliation{MTA-DE Particle Physics Research Group, P.O.Box 51,
H-4001 Debrecen, Hungary}

\author{K. Sailer}
\affiliation{Department of Theoretical Physics, University of Debrecen,
P.O. Box 5, H-4010 Debrecen, Hungary}

\begin{abstract}

The energy gap between the first excited state and the ground state is calculated for
the quantized anharmonic oscillator in the framework of the functional renormalization group 
method. The compactly supported smooth regulator is used which includes various types of
regulators as limiting cases. It was found that the value of the energy gap depends on the
regulator parameters. We argue that the optimization based on the disappearance of the false,
broken symmetric phase of the model leads to the Litim's regulator. The least sensitivity on the
regulator parameters leads however to  an IR regulator being somewhat different of
the Litim's one, but it can be described as a perturbatively improved, or generalized Litim's
regulator and provides analytic evolution equations, too.

\end{abstract}

\pacs{11.10.Gh,11.10.Hi,05.10.Cc}
\keywords{renormalization group, critical exponents}
\maketitle

\section{Introduction}

The application of the functional renormalization group (RG) method
\cite{Wetterich:1992yh,Berges:2000ew,Polonyi:2001se,Pawlowski:2005xe,Gies:2006wv,Rosten:2010vm}
for the quantized 1-dimensional anharmonic oscillator is a highly nontrivial task
\cite{Kapoyannis:2000sp,Zappala:2001nv,Nagy:2010fv}.
In the numerical treatment of the RG the problem  arises from the fact that the ultraviolet (UV) double-well potential
cannot become convex in the infrared (IR) regime, if the coupling of the anharmonic
term is weak. The potential of the classical model, i.e., the potential at the UV scale
can be either a simple convex potential or a double-well potential with  non-trivial minima.
In the second case the classical model has ground states with spontaneously broken $Z_2$ symmetry.
In quantum mechanics, however, the effective potential should be convex due to the tunneling 
effect even if the RG evolution is  started from the double-well UV potential; so the 
quantized model should have a single symmetric phase.

Usually the energy gap between the ground state and the first excited state is
the observable calculated for the anharmonic oscillator. The model can be also handled by
solving the Schr\"odinger equation for the anharmonic potential numerically. Let us call
the  results of the latter exact. The exact results can then serve as guidelines in
controlling and optimizing the results of the RG calculations. 

The original optimization strategy is based on determining the IR regulator which provides the
fastest convergence for the RG flows \cite{Litim:2000ci,Litim:2001up}.
A plausible optimization condition could be to find the IR regulator which gives the closest
to the exact value of the observable. The problem of this reasoning is that, on the one hand, 
we have numerical results in RG with strong truncations in the gradient expansion, and the 
Taylor expansions in its functionals. On the other hand, the regulators can be deformed easily
to provide a wide range of results which may reproduce any exact values for an observable.
Therefore, we should choose another strategy for the optimization of the IR regulator.
In this paper we follow the optimization strategy which is based on choosing that value of
the observable as the optimized one that shows the slightest dependence on the parameters of the
regulator. This optimization strategy has often been used recently
\cite{Canet:2002gs,Canet:2003qd,Essafi:2012hr,Li:2012fe,Li:2013mda,Essafi:2014dla,Kloss:2013xva}.
At first glance it seems to be rather a mathematical condition. Nevertheless,
it can be made plausible by simple physical reasoning. The original 
generating functional does not contain the IR regulator, therefore it is reasonable
to look for such physical results which have the least dependence
on the regulator. However the systematic search for the extremal value
of any observable could not be performed among the various available IR regulators due 
to their rather different functional forms. This situation has been changed as
the compactly supported smooth (css) IR regulator function has been introduced
\cite{Nandori:2012tc} inspired by the so-called Salamon-Vertse potential used in nuclear physics
\cite{Salamon:2008zz,Salamon:2010bg,Racz:2011dz,Salamon:2012pw}.
One can recast the css regulator into a simpler form that enables one to deform 
it continuously into the Litim's, the exponential and the power-law regulators by using
only two parameters \cite{Nagy:2013hka}, and to perform the optimization
program on a simple 2-dimensional surface. By this technique we found successfully the
least sensitive extremal value of the critical exponent
$\nu$ of the correlation length for the quantum Einstein gravity and for the 3-dimensional
$O(1)$ model \cite{Nagy:2013hka}. Then, this method was also used successfully to 
investigate sine-Gordon type models \cite{Marian:2013zza}. In the present
paper the same approach is used for the determination of the energy gap of the
1-dimensional anharmonic oscillator.

We can find an optimized regulator different of the Litim's optimized one. However,
the obtained optimized css regulator is very close to the Litim's result,
so that one can consider it the perturbative generalization of the Litim's regulator.
For the resulting new regulator one can perform the momentum integration in
the RG equation in order to obtain for it a closed analytic form and that makes
much more simple to handle the RG equation numerically.

The paper is organized as follows. In \sect{sec:evol} the investigated model, the RG method,
and the regulators are introduced. In \sect{sec:trunc} we discuss the truncations applied
during the numerical calculations. We collect the results for the optimization strategies
in \sect{sec:opt}. Finally, in \sect{sec:sum} the conclusions are drawn up.

\section{Evolution equations}\label{sec:evol}

The RG method provides us a partial integro-differential equation for the effective action,
which is called the Wetterich equation \cite{Wetterich:1992yh,Berges:2000ew}
\beq\label{WRG}
\dot\Gamma_k=\frac12\mbox{Tr}\frac{\dot R_k}{R_k+\Gamma''_k},
\eeq
where $^.=k\partial_k$, $^\prime=\partial/\partial\phi$, $R_k$ is the regulator and the trace Tr denotes the
integration over all momenta and summation for internal indices.
\eqn{WRG} has been solved over the functional subspace defined by the ansatz
\beq\label{eaans}
\Gamma_k = \int_x\left[\frac{Z_k}2 (\partial_\mu\phi)^2+V_k\right],
\eeq
with the potential $V_k$, and the wave function renormalization $Z_k$. In the case of
the local potential approximation (LPA) $Z_k=1$.
Quantum mechanics can be considered as a quantum field theory with 0 spatial and 1 time
dimension, therefore one can apply the RG technique there, the field variable $\phi$
represents  the oscillator coordinate. Then the evolution equation for the potential reads as
\beq\label{potRG}
\dot V_k=\frac {1}{2\pi}\int_0^{\infty} dp \frac {\dot R_k}{p^2+R_k+V''_k},
\eeq
where $p$ stands essentially for the frequency in that case. The initial condition for \eqn{WRG}
is given by the explicit form of the microscopic effective action at the ultra violet (UV)
cutoff $k=\Lambda$. There are lots of examples in the literature for different types of regulator functions.
Here we use the dimensionless form of the css regulator,
\beq
\label{rcss}
r_{css} = \frac{s_1}{\exp[s_1 y^b/(1 -s_2 y^b)] -1} \theta(1-s_2 y^b),
\eeq
with $y=p^2/k^2$ and $r=r(y)$ is the dimensionless regulator $r=R/p^2$, furthermore
$b\ge 1$ and $s_1$, $s_2$ are positive parameters. Unfortunately, the momentum integral
in the evolution equation \eqn{potRG} has no analytic form for this regulator.
For the limiting cases of the css regulator one recovers the following commonly used regulator functions \cite{Nagy:2013hka},
\bea
\lim_{s_1\to 0} r_{css} &=& \left(\frac{1}{y^b} -s_2\right) \theta(1-s_2 y^b),\label{limlit}\\
\lim_{s_1\to 0,s_2\to 0} r_{css} &=& \frac1{y^b},\label{limpow}\\
\lim_{s_2\to 0} r_{css} &=& \frac{s_1}{\exp[s_1 y^b]-1}\label{limexp}.
\eea
where the first limit gives the Litim's regulator for $s_2=1$, the second one is the
power-law regulator, and the third one gives the exponential regulator, if $s_1=1$.
One can perform the optimization by finding an extremum of the energy gap on the
parameter space spanned by $s_1$ and $s_2$. We note that the case $b=1$ satisfies the
normalization conditions \cite{Reuter:2001ag}
\beq\label{reglim}
\lim_{y\to 0} yr = 1 ~\mbox{and}~\lim_{y\to\infty} yr = 0.
\eeq
The usage of the power-law regulator with $b=1$ is usually called Callan-Symanzik (CS) scheme.
We investigate the quantum mechanical anharmonic oscillator in terms of the Taylor-expanded potential
\beq\label{pot}
V_k=\frac {m_k^2}{2} \phi ^{2}+g_k \phi ^{4}+\sum _{n=3}^N \frac {g_{2n}(k)}{(2n)!} \phi ^{2n},
\eeq
where, besides the harmonic and the quartic anharmonic terms we have introduced the 
additional couplings $g_{2n}$ with $n\ge 3$ which are generated by the RG method.
Inserting \eqn{pot} into \eqn{potRG} we obtain a system of ordinary differential
equations for the couplings, as usual. The evolution equations for the couplings $m_k$
and $g_k$ are:
\bea
\dot m^2_k&=& - \frac{12}{\pi}\int_0^{\infty} dp \dot R_k\frac{g_k}{(p^2+R_k+m^2_k)^2},\nn
\dot g_k&=&\frac {1}{48\pi}\int_0^{\infty} dp \dot R_k\left[\frac{3456g_k^2}{(p^2+R_k+m^2_k)^3}
-\frac{g_6}{(p^2+R_k+m^2_k)^2}\right].
\eea
The evolution equations for the further couplings have similar qualitative structures.

The solution of the RG equations in LPA provides us the effective potential $V_0$, i.e., the potential in the 
limit $k \to 0$. We look for the energy gap of the model given by
\beq
\Delta E = \sqrt{ V_0^{\prime\prime} }\bigg |_{\phi=<\phi>},
\eeq
where $\langle \phi\rangle$ is the vacuum expectation value of the field variable.
In quantum mechanics the vacuum expectation value is the trivial field configuration,
i.e., $\langle \phi \rangle = 0$. From \eqn{pot} the energy gap is
\beq
\Delta E = m_0,
\eeq
which is the IR limit of the coupling $m_k$.

\section{Truncations}\label{sec:trunc}

Performing the RG analysis of the quantized 1-dimensional anharmonic oscillator
we have used two kinds of truncations, that of the gradient expansion in its lowest order,
the LPA, and that of the Taylor-expansion of the local potential.
We have restricted ourselves to the LPA because the field-dependence of the wave-function
renormalization cannot be handled by Taylor expansion due to its strange functional form
\cite{Zappala:2001nv}. At the UV scale chosen for $\Lambda=1500$ we set the couplings
$m_\Lambda^2$ and $g_\Lambda$ and the further couplings are suppressed. We investigate the energy
gap $\Delta E$ as the function of these initial values.

\begin{figure}
\centerline{ \includegraphics[width=6cm,angle=-90]{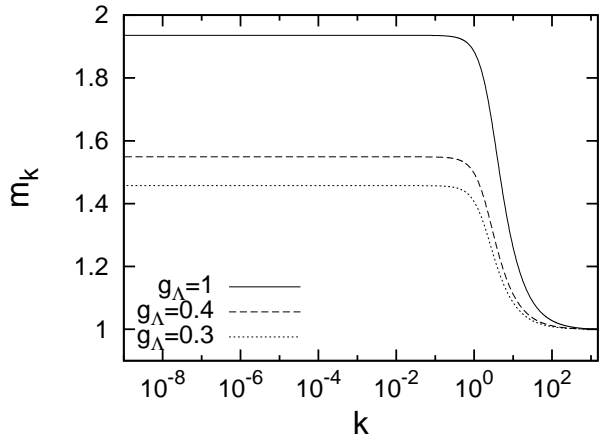}}
\caption{\label{fig:fig1} The evolution of the coupling $m_k$ is shown for Callan-Symanzik
scheme, for  $m^2_\Lambda=1$. The  curves correspond to various initial values of $g_\Lambda$.}
\end{figure}

\fig{fig:fig1} shows the flow of the coupling $m_k$ during the evolution in CS scheme
for various initial values of the quartic coupling $g_\Lambda$.
In the IR limit the dimensionful coupling $m_k^2$ as well as the other dimensionful 
couplings scale marginally, i.e., they tend to positive constant values.

The obtained numerical value of the energy gap is sensitive to  many parameters in the
calculations. Ideally one should optimize the values of the
energy gap for the regulator parameters $b$, $s_1$, $s_2$, and for $N$, i.e., the order of the expansion in \eqn{pot}.
Throughout the present work we set $b=1$ because only this choice satisfies the normalization
condition in \eqn{reglim} for the regulator. We note, on the one hand, that previous results in the literature
showed that the optimal value is around $b\approx 2$ in the 2-dimensional sine-Gordon model
\cite{Nandori:2010ij} and in the 3-dimensional  $O(N)$ model.
On the other hand, it is impossible to find an optimal value for the energy gap of the 1-dimensional
anharmonic oscillator by varying the value of the parameter $b$, because for various initial values of
$m_\Lambda^2$ and $g_\Lambda$ one obtains various `optimal' values in the interval
$b\in(1\ldots 6)$ \cite{Kovacs:2013acta}.

\begin{figure}
\centerline{ \includegraphics[width=6cm,angle=-90]{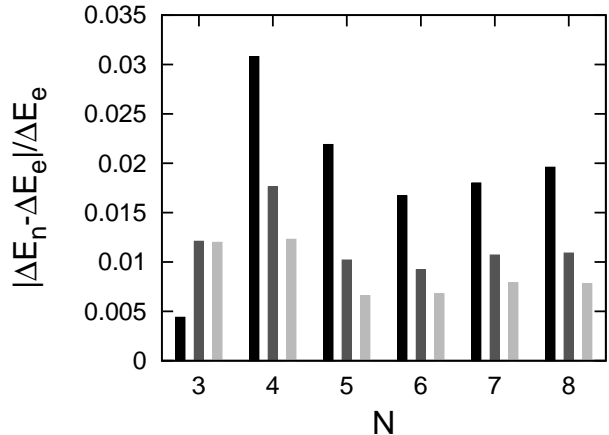}}
\caption{\label{fig:fig2} The relative deviation of the numerical values of
the energy gap ($\Delta E_n$) from the exact ones ($\Delta E_e$) as the function of $N$ is shown
for various initial values of $g_\Lambda$ for $m^2_\Lambda=-1$. The black column corresponds to $g_\Lambda=0.4$, the
dark grey column denotes $g_\Lambda=0.3$ and the light grey one refers to $g_\Lambda=0.2$. The data were calculated in
the CS scheme.}
\end{figure}

The power-law regulator was chosen to explore the $N$-dependence, the results are demonstrated in \fig{fig:fig2}.
We choose the case $m^2_\Lambda<0$ for the optimization. Although one expects that larger
values of $N$ could improve the approximation of the expansion, one can see in \fig{fig:fig2}
that the numerical errors increase for too large $N$ values. Thus one concludes that
the optimal value of the number of couplings is about 6 in the LPA for the power-law regulator. 

The results for the most important regulators are collected in \tab{tab:tab1}.
\begin{table}
\begin{center}
\begin{tabular}{|c|c|c|c|c|c|c|}
\hline
$m^2_\Lambda$ &$g_\Lambda$ & $\Delta E_{exact}$ & $\Delta E_{HK}$ & $\Delta E_{Litim}$ & $\Delta E_{CS}$ & $\Delta E_{exp}$ \nn \hline
1 & 1 &	1.9341 & 1.9380	& 1.9386 & 1.9358 & 1.9382 \nn  \hline
1 & 0.4& 1.5482 & 1.5498 & 1.5507 & 1.5490 & 1.5504 \nn \hline
1 & 0.1 & 1.2104 & 1.2109 & 1.2110 & 1.2105 & 1.2109 \nn \hline
1 & 0.05 & 1.1208 & 1.1210 & 1.1211 & 1.1208& 1.1210 \nn \hline
1 & 0.03 & 1.0779 & 1.0780 & 1.0780 & 1.0779 & 1.0778 \nn \hline
1 & 0.02 & 1.0540 & 1.0542 & 1.0542 & 1.0541 & 1.0542 \nn \hline
-1 & 0.4 & 0.9667 & 0.9730 & 0.9778 & 0.9733 & 0.9772 \nn \hline
-1 & 0.3 & 0.8166 & 0.8233 & 0.8288 & 0.8241 & 0.8281 \nn  \hline
-1 & 0.2 & 0.6159 & 0.6227 & 0.6309 & 0.6262 & 0.6302 \nn \hline
\end{tabular}
\caption{\label{tab:tab1} The value of the first energy gap is shown for
various initial conditions. In the order of the columns it is shown
the well-known exact values, the values calculated by Heat Kernel renormalization
and the values calculated by us in the Litim, Callan-Symanzik and  exponential schemes.}
\end{center}
\end{table}
For negative values of $m_\Lambda^2$ and for small values of $g_\Lambda$ the RG approach does not work,
the effective potential becomes concave at $\phi=0$ and we have no result for the energy gap.
This happens presumably due to the strong truncation of the potential and that of the gradient expansion.
Table \ref{tab:tab1} shows that the choice $N=6$ provides rather close values of the energy
gap for the various kinds of regulators. Their deviations 
from each other are much less then their deviations of $\sim 1$ \% from the exact values for $m_\Lambda^2>0$.
Similar is true when $m_\Lambda^2<0$ and when $g_\Lambda$ is sufficiently large.

\section{Optimization strategies}\label{sec:opt}

\begin{center}
\begin{figure}
\includegraphics[width=7cm,angle=-90]{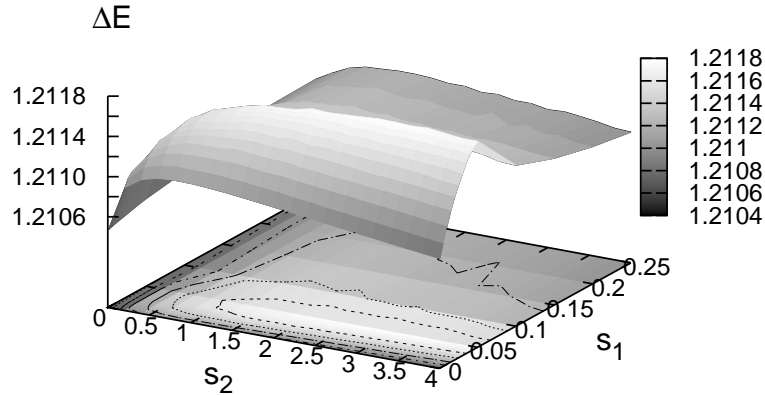}
\caption{\label{fig:sur} The energy gap $\Delta E$ is shown as the function of the regulator parameters $s_1$ and $s_2$.
The initial couplings are $m^2_\Lambda=1$ and $g_\Lambda=0.1$.
We set $b=1$.}
\end{figure}
\end{center}

For the further investigations we set $N=6$ and look for the extremum of the energy gap in the parameters $s_1$ and $s_2$.
In \fig{fig:sur} we plotted $\Delta E$ for different regulator parameters and for positive 
$m_\Lambda^2$.
Interestingly the results show very slight regulator-dependence. We had to go beyond 3 digits in the numerical
precision to find some nontrivial results.
As a comparison we note that in the case of the optimization of the model of quantum Einstein gravity \cite{Nagy:2013hka}
we obtained very strong regulator-dependence. There the value of the critical exponent could change
several orders of magnitude, and even its sign could change.
The anharmonic oscillator investigated here is a 1-dimensional model and this can 
be the reason of such a weak  regulator-dependence.
The removal of UV divergences in higher dimensional models may introduce strong 
scheme-dependence, while there is no need to remove UV divergences in 1-dimensional models.
Furthermore there is no IR singularity due to the positive mass term.

It would be the most straightforward optimization strategy to recover the exact, i.e., physical value of 
the observable $\Delta E$ for a given IR regulator.
Fig. \ref{fig:sur} shows for a particular choice of the initial conditions,
that the exact value of the energy gap $\Delta E=1.2104$ can be obtained by the 
power-law regulator near the origin $s_1\approx0$ and $s_2\approx 0$ of the parameter space. 
Unfortunately, other initial conditions require other IR regulators with different
regulator parameters. Although this strategy can be supported mostly
by physical arguments and it can be the only reasonable optimization, 
nevertheless it does not work in our RG framework. This strategy may work
when the truncations are minimal, which is not the case in our treatment.

\begin{figure}
\centerline{ \includegraphics[width=6cm,angle=-90]{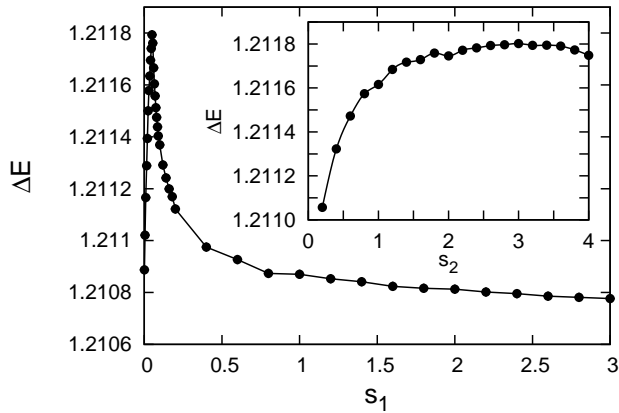}}
\caption{\label{fig:sect} The sections of the surface plot are shown through the extremum.
The initial couplings are $m^2_\Lambda=1$ and $g_\Lambda=0.05$. We set $s_2=3$ and
$s_1=0.05$ in the inset.}
\end{figure}

It is another possibility for the optimization to look for that value of the observable
$\Delta E$ which shows the least sensitivity to the regulator parameters, i.e., to find an
extremum of the energy gap in the parameter space. At the maximum of the surface in \fig{fig:sur}
the sensitivity of $\Delta E$ is minimal to $s_1$ and $s_2$. Fig. \ref{fig:sur} shows, that there
is a maximum of the energy gap at $s_1=0.05$ and $s_2=3$.
Accordingly the optimized regulator corresponds to the css regulator of the form
\beq\label{regopt}
r_{opt} = \frac{0.05}{\exp[0.05 y/(1 -3 y)] -1} \theta(1-3 y).
\eeq
For other initial couplings we have got by means of this least-sensitivity optimization the same
results with the same optimized IR regulator. The value of $s_1=0$ corresponds to the limit
of a general Litim's regulator in \eqn{limlit}. \fig{fig:sur} shows that this Litim's
regulator zone gives neither the minimal value nor an extremum for $\Delta E$.
In \fig{fig:sect} we demonstrate how the energy gap changes close to the Litim's limit.
There is a maximum at $s_1=0.05$ for practically all of the sections with $s_2=$const.
These maxima create a saddle along the $s_2$ direction for the small values $s_1\approx 0.05$.
The $s_1=0.05$ section of  the saddle is plotted in the inset of
\fig{fig:sect}. This curve also has a local maximum at $s_2=3$.

The minimal sensitivity optimization works well if we choose a positive initial value  $m^2_\Lambda$
when both the blocked potential and the resulting effective potential are  convex.
However, for negative initial values  $m^2_\Lambda$ the convexity 
cannot be granted. The quantized anharmonic oscillator can have only
a symmetric phase, the spontaneously broken symmetric phase is excluded by the tunneling effect.
For sufficiently small initial values of $g_\Lambda$ there is no room to turn the concave blocked potential
into a convex one during the RG evolution. This appears probably due to the strongly truncated
gradient expansion of the effective action \cite{Zappala:2001nv,Kapoyannis:2000sp}.
Taking into account the wave-function renormalization and solving the RG equations without
Taylor expansion may improve the treatment, i.e., may enable one to determine
$\Delta E$ even for smaller initial values of $g_\Lambda$.

The determination of the smallest initial value of $g_\Lambda$ for which the effective potential becomes
convex gives another possibility to optimize the IR regulator, since the model should have only a symmetric phase.
We note that the disappearance of the false phase has been used recently to find optimized IR regulators
for sine-Gordon type models \cite{Nandori:2013nda}. We found that $g_\Lambda=0.08$ is the smallest
initial value for which the energy gap can be determined reliably in the RG framework used by us.
We note that in \cite{Zappala:2001nv} the smallest value is $g_\Lambda=0.02$ which is a better result. Here we
cannot have such a precision, because we did not include the wave-function renormalization
and  Taylor-expanded the potential that was avoided in \cite{Zappala:2001nv}.
Nevertheless, the issue of optimization is important even if the RG framework involves quite strong truncations,
like in our case. In \tab{tab:tab2} we collected the results for the energy gap for $m^2_\Lambda=-1$ and $g_\Lambda=0.08$.
\begin{table}
\begin{center}
\begin{tabular}{|c|c|c|c|c|}
\hline
 & $s_2=0.001$ & $s_2=1$ & $s_2=2$ & $s_2=3$ \nn \hline
$s_1=0.001$ & -- & 0.23538 & 0.23679 & 753913.25671 \nn \hline
$s_1=0.05$ & -- & 0.23598 & 0.23785 & 23102.53408 \nn \hline
$s_1=1$ & 0.23766 & 0.23556 & 341061.15077 & -- \nn \hline
$s_1=2$ & 0.23612 & 0.23604 & -- & -- \nn \hline
$s_1=3$ & 0.23566 & 0.23644 & -- & -- \nn \hline
\end{tabular}
\caption{\label{tab:tab2} The energy gap is shown as the function of the regulator parameters
$s_1$ and $s_2$ for the initial couplings $m^2_\Lambda=-1$ and $g_\Lambda=0.08$.}
\end{center}
\end{table}
The extremely large values show numerical instabilities during the calculations.
\tab{tab:tab2} shows that the IR regulator in \eqn{regopt} is not the optimized one in \eqn{regopt}.
If one defines the optimized regulator via finding the smallest $g_\Lambda$ which restores
the convexity of the potential, then the Litim's regulator proves to be the best one,
since it gives the smallest value of $\Delta E$ there.
It seems that various optimization procedures give different IR regulators.

The IR regulator \eqn{regopt} is very close to the Litim's one.
If one Taylor expands  the css regulator in $s_1$ at $s_1=0$ then one obtains that
\beq\label{litimgen}
r_{pert}\approx\left(\frac {1 -s_2 y^b}{y^b}-\frac{s_1}{2}+\frac{y^bs_1^2}{12(1 -s_2 y^b)}+\ldots\right)
\theta(1-s_2 y^b).
\eeq
Up to the linear term in $s_1$ the LPA evolution equation for the potential takes the form
\beq
\dot V = \frac {1}{2\pi}\int_0^{\frac{k}{\sqrt{s_2}}} dp \frac {2k^2}{k^2 +p^2(1-s_2-\frac{s_1}{2})+V''}.
\eeq
The perturbative Litim's regulator in \eqn{litimgen} does not cancel the momentum dependence in the
integrand of the loop integral. In this sense the regulator $r_{pert}$ takes after the 
CS type regulator, however the UV divergence does not appear, since the upper
integration limit is restricted by the $\theta$ function and it guarantees the finiteness in any
dimensions. Moreover, the resulting RG equation remains analytic. In $d=1$ it reads as
\beq
\dot V = \frac {k^2}{\pi}\frac1{\sqrt{(1-s_2-\frac{s_1}{2})(k^2+V'')}}
\tan^{-1}\left(\sqrt{\frac{(1-s_2-\frac{s_1}{2})k^2}{s_2(k^2+V'')}}\right).
\eeq

\section{Conclusions}\label{sec:sum}

By using the functional renormalization group method we calculated the energy gap for the quantized
1-dimensional anharmonic oscillator. The renormalization group approach requires approximations
 which can introduce some regulator-dependence. We used the local potential approximation and the 
Taylor expansion of the potential with a truncation yielding the smallest deviation of the
energy gap of the oscillator from its exact value. The regulator-dependence of the results has been
investigated by making use of the css regulator that enables one to consider various types of regulator
functions in a unique parametrization. The css regulator \eqn{rcss} depends on the parameters $b$, $s_1$ and $s_2$.
We set $b=1$ for our study when the normalization conditions \eqn{reglim} are satisfied. The optimization of the css regulator
with respect to the parameters  $s_1$ and $s_2$ has been carried out.

For the anharmonic oscillator with a single-well UV potential, it turned out that the optimization strategy based on
the minimal sensitivity on the regulator parameters works rather well. It is found
that  the energy gap has an extremum as the function of the regulator parameters. The optimized
regulator found in that manner is shown to be the generalization of the Litim's regulator and it
provides an analytic evolution equation for the potential in $d=1$. For the anharmonic oscillator
with a double-well UV potential, this generalized Litim's regulator seems not to be the optimal one.
Instead of the optimization via achieving the minimal sensitivity of the observable on the regulator parameters
another optimization strategy can be followed. Then one looks for the regulator that enables one
to reestablish the convexity of the numerically determined effective potential for the smallest value of
the quartic coupling. We have found that the Litim's regulator appears in that case rather optimal instead of 
the generalized Litim's regulator introduced in the case of the single-well potential. It is argued that
such a situation is due to the strong truncations in the gradient expansion and in the Taylor-expansion of the potential.

\section*{Acknowledgments}

This research was supported by the European Union and the State of Hungary, co-financed by the
European Social Fund in the framework of T\'AMOP-4.2.4.A/2-11/1-2012-0001 'National Excellence Program'.
(Author: S\'andor Nagy)

\bibliography{nagy}

\begin{thebibliography}{29}
\expandafter\ifx\csname natexlab\endcsname\relax\def\natexlab#1{#1}\fi
\expandafter\ifx\csname bibnamefont\endcsname\relax
  \def\bibnamefont#1{#1}\fi
\expandafter\ifx\csname bibfnamefont\endcsname\relax
  \def\bibfnamefont#1{#1}\fi
\expandafter\ifx\csname citenamefont\endcsname\relax
  \def\citenamefont#1{#1}\fi
\expandafter\ifx\csname url\endcsname\relax
  \def\url#1{\texttt{#1}}\fi
\expandafter\ifx\csname urlprefix\endcsname\relax\def\urlprefix{URL }\fi
\providecommand{\bibinfo}[2]{#2}
\providecommand{\eprint}[2][]{\url{#2}}

\bibitem[{\citenamefont{Wetterich}(1993)}]{Wetterich:1992yh}
\bibinfo{author}{\bibfnamefont{C.}~\bibnamefont{Wetterich}},
  \bibinfo{journal}{Phys.Lett.} \textbf{\bibinfo{volume}{B301}},
  \bibinfo{pages}{90} (\bibinfo{year}{1993}).

\bibitem[{\citenamefont{Berges et~al.}(2002)\citenamefont{Berges, Tetradis, and
  Wetterich}}]{Berges:2000ew}
\bibinfo{author}{\bibfnamefont{J.}~\bibnamefont{Berges}},
  \bibinfo{author}{\bibfnamefont{N.}~\bibnamefont{Tetradis}}, \bibnamefont{and}
  \bibinfo{author}{\bibfnamefont{C.}~\bibnamefont{Wetterich}},
  \bibinfo{journal}{Phys.Rept.} \textbf{\bibinfo{volume}{363}},
  \bibinfo{pages}{223} (\bibinfo{year}{2002}), \eprint{hep-ph/0005122}.

\bibitem[{\citenamefont{Polonyi}(2003)}]{Polonyi:2001se}
\bibinfo{author}{\bibfnamefont{J.}~\bibnamefont{Polonyi}},
  \bibinfo{journal}{Central Eur.J.Phys.} \textbf{\bibinfo{volume}{1}},
  \bibinfo{pages}{1} (\bibinfo{year}{2003}), \eprint{hep-th/0110026}.

\bibitem[{\citenamefont{Pawlowski}(2007)}]{Pawlowski:2005xe}
\bibinfo{author}{\bibfnamefont{J.~M.} \bibnamefont{Pawlowski}},
  \bibinfo{journal}{Annals Phys.} \textbf{\bibinfo{volume}{322}},
  \bibinfo{pages}{2831} (\bibinfo{year}{2007}), \eprint{hep-th/0512261}.

\bibitem[{\citenamefont{Gies}(2012)}]{Gies:2006wv}
\bibinfo{author}{\bibfnamefont{H.}~\bibnamefont{Gies}},
  \bibinfo{journal}{Lect.Notes Phys.} \textbf{\bibinfo{volume}{852}},
  \bibinfo{pages}{287} (\bibinfo{year}{2012}), \eprint{hep-ph/0611146}.

\bibitem[{\citenamefont{Rosten}(2012)}]{Rosten:2010vm}
\bibinfo{author}{\bibfnamefont{O.~J.} \bibnamefont{Rosten}},
  \bibinfo{journal}{Phys.Rept.} \textbf{\bibinfo{volume}{511}},
  \bibinfo{pages}{177} (\bibinfo{year}{2012}), \eprint{1003.1366}.

\bibitem[{\citenamefont{Kapoyannis and Tetradis}(2000)}]{Kapoyannis:2000sp}
\bibinfo{author}{\bibfnamefont{A.}~\bibnamefont{Kapoyannis}} \bibnamefont{and}
  \bibinfo{author}{\bibfnamefont{N.}~\bibnamefont{Tetradis}},
  \bibinfo{journal}{Phys.Lett.} \textbf{\bibinfo{volume}{A276}},
  \bibinfo{pages}{225} (\bibinfo{year}{2000}), \eprint{hep-th/0010180}.

\bibitem[{\citenamefont{Zappala}(2001)}]{Zappala:2001nv}
\bibinfo{author}{\bibfnamefont{D.}~\bibnamefont{Zappala}},
  \bibinfo{journal}{Phys.Lett.} \textbf{\bibinfo{volume}{A290}},
  \bibinfo{pages}{35} (\bibinfo{year}{2001}), \eprint{quant-ph/0108019}.

\bibitem[{\citenamefont{Nagy and Sailer}(2011)}]{Nagy:2010fv}
\bibinfo{author}{\bibfnamefont{S.}~\bibnamefont{Nagy}} \bibnamefont{and}
  \bibinfo{author}{\bibfnamefont{K.}~\bibnamefont{Sailer}},
  \bibinfo{journal}{Annals Phys.} \textbf{\bibinfo{volume}{326}},
  \bibinfo{pages}{1839} (\bibinfo{year}{2011}), \eprint{1009.4041}.

\bibitem[{\citenamefont{Litim}(2000)}]{Litim:2000ci}
\bibinfo{author}{\bibfnamefont{D.~F.} \bibnamefont{Litim}},
  \bibinfo{journal}{Phys.Lett.} \textbf{\bibinfo{volume}{B486}},
  \bibinfo{pages}{92} (\bibinfo{year}{2000}), \eprint{hep-th/0005245}.

\bibitem[{\citenamefont{Litim}(2001)}]{Litim:2001up}
\bibinfo{author}{\bibfnamefont{D.~F.} \bibnamefont{Litim}},
  \bibinfo{journal}{Phys.Rev.} \textbf{\bibinfo{volume}{D64}},
  \bibinfo{pages}{105007} (\bibinfo{year}{2001}), \eprint{hep-th/0103195}.

\bibitem[{\citenamefont{Canet et~al.}(2003{\natexlab{a}})\citenamefont{Canet,
  Delamotte, Mouhanna, and Vidal}}]{Canet:2002gs}
\bibinfo{author}{\bibfnamefont{L.}~\bibnamefont{Canet}},
  \bibinfo{author}{\bibfnamefont{B.}~\bibnamefont{Delamotte}},
  \bibinfo{author}{\bibfnamefont{D.}~\bibnamefont{Mouhanna}}, \bibnamefont{and}
  \bibinfo{author}{\bibfnamefont{J.}~\bibnamefont{Vidal}},
  \bibinfo{journal}{Phys.Rev.} \textbf{\bibinfo{volume}{D67}},
  \bibinfo{pages}{065004} (\bibinfo{year}{2003}{\natexlab{a}}),
  \eprint{hep-th/0211055}.

\bibitem[{\citenamefont{Canet et~al.}(2003{\natexlab{b}})\citenamefont{Canet,
  Delamotte, Mouhanna, and Vidal}}]{Canet:2003qd}
\bibinfo{author}{\bibfnamefont{L.}~\bibnamefont{Canet}},
  \bibinfo{author}{\bibfnamefont{B.}~\bibnamefont{Delamotte}},
  \bibinfo{author}{\bibfnamefont{D.}~\bibnamefont{Mouhanna}}, \bibnamefont{and}
  \bibinfo{author}{\bibfnamefont{J.}~\bibnamefont{Vidal}},
  \bibinfo{journal}{Phys.Rev.} \textbf{\bibinfo{volume}{B68}},
  \bibinfo{pages}{064421} (\bibinfo{year}{2003}{\natexlab{b}}),
  \eprint{hep-th/0302227}.

\bibitem[{\citenamefont{Essafi et~al.}(2012)\citenamefont{Essafi, Kownacki, and
  Mouhanna}}]{Essafi:2012hr}
\bibinfo{author}{\bibfnamefont{K.}~\bibnamefont{Essafi}},
  \bibinfo{author}{\bibfnamefont{J.}~\bibnamefont{Kownacki}}, \bibnamefont{and}
  \bibinfo{author}{\bibfnamefont{D.}~\bibnamefont{Mouhanna}},
  \bibinfo{journal}{Europhys.Lett.} \textbf{\bibinfo{volume}{98}},
  \bibinfo{pages}{51002} (\bibinfo{year}{2012}), \eprint{1202.5946}.

\bibitem[{\citenamefont{Li and Luo}(2012)}]{Li:2012fe}
\bibinfo{author}{\bibfnamefont{M.-F.} \bibnamefont{Li}} \bibnamefont{and}
  \bibinfo{author}{\bibfnamefont{M.}~\bibnamefont{Luo}},
  \bibinfo{journal}{Phys.Rev.} \textbf{\bibinfo{volume}{D85}},
  \bibinfo{pages}{085027} (\bibinfo{year}{2012}), \eprint{1201.2468}.

\bibitem[{\citenamefont{Li and Luo}(2013)}]{Li:2013mda}
\bibinfo{author}{\bibfnamefont{M.-F.} \bibnamefont{Li}} \bibnamefont{and}
  \bibinfo{author}{\bibfnamefont{M.}~\bibnamefont{Luo}},
  \bibinfo{journal}{Phys.Rev.} \textbf{\bibinfo{volume}{D88}},
  \bibinfo{pages}{085019} (\bibinfo{year}{2013}), \eprint{1305.2578}.

\bibitem[{\citenamefont{Essafi et~al.}(2014)\citenamefont{Essafi, Kownacki, and
  Mouhanna}}]{Essafi:2014dla}
\bibinfo{author}{\bibfnamefont{K.}~\bibnamefont{Essafi}},
  \bibinfo{author}{\bibfnamefont{J.~P.} \bibnamefont{Kownacki}},
  \bibnamefont{and} \bibinfo{author}{\bibfnamefont{D.}~\bibnamefont{Mouhanna}}
  (\bibinfo{year}{2014}), \eprint{1402.0426}.

\bibitem[{\citenamefont{Kloss et~al.}(2014)\citenamefont{Kloss, Canet,
  Delamotte, and Wschebor}}]{Kloss:2013xva}
\bibinfo{author}{\bibfnamefont{T.}~\bibnamefont{Kloss}},
  \bibinfo{author}{\bibfnamefont{L.}~\bibnamefont{Canet}},
  \bibinfo{author}{\bibfnamefont{B.}~\bibnamefont{Delamotte}},
  \bibnamefont{and} \bibinfo{author}{\bibfnamefont{N.}~\bibnamefont{Wschebor}},
  \bibinfo{journal}{Phys.Rev.} \textbf{\bibinfo{volume}{E89}},
  \bibinfo{pages}{022108} (\bibinfo{year}{2014}), \eprint{1312.6028}.

\bibitem[{\citenamefont{Nandori}(2013)}]{Nandori:2012tc}
\bibinfo{author}{\bibfnamefont{I.}~\bibnamefont{Nandori}},
  \bibinfo{journal}{JHEP} \textbf{\bibinfo{volume}{1304}}, \bibinfo{pages}{150}
  (\bibinfo{year}{2013}), \eprint{1208.5021}.

\bibitem[{\citenamefont{Salamon and Vertse}(2008)}]{Salamon:2008zz}
\bibinfo{author}{\bibfnamefont{P.}~\bibnamefont{Salamon}} \bibnamefont{and}
  \bibinfo{author}{\bibfnamefont{T.}~\bibnamefont{Vertse}},
  \bibinfo{journal}{Phys.Rev.} \textbf{\bibinfo{volume}{C77}},
  \bibinfo{pages}{037302} (\bibinfo{year}{2008}).

\bibitem[{\citenamefont{Salamon et~al.}(2010)\citenamefont{Salamon, Kruppa, and
  Vertse}}]{Salamon:2010bg}
\bibinfo{author}{\bibfnamefont{P.}~\bibnamefont{Salamon}},
  \bibinfo{author}{\bibfnamefont{A.}~\bibnamefont{Kruppa}}, \bibnamefont{and}
  \bibinfo{author}{\bibfnamefont{T.}~\bibnamefont{Vertse}},
  \bibinfo{journal}{Phys.Rev.} \textbf{\bibinfo{volume}{C81}},
  \bibinfo{pages}{064322} (\bibinfo{year}{2010}), \eprint{1002.4333}.

\bibitem[{\citenamefont{Racz et~al.}(2011)\citenamefont{Racz, Salamon, and
  Vertse}}]{Racz:2011dz}
\bibinfo{author}{\bibfnamefont{A.}~\bibnamefont{Racz}},
  \bibinfo{author}{\bibfnamefont{P.}~\bibnamefont{Salamon}}, \bibnamefont{and}
  \bibinfo{author}{\bibfnamefont{T.}~\bibnamefont{Vertse}},
  \bibinfo{journal}{Phys.Rev.} \textbf{\bibinfo{volume}{C84}},
  \bibinfo{pages}{037602} (\bibinfo{year}{2011}), \eprint{1107.2217}.

\bibitem[{\citenamefont{Salamon et~al.}(2012)\citenamefont{Salamon, Vertse, and
  Balkay}}]{Salamon:2012pw}
\bibinfo{author}{\bibfnamefont{P.}~\bibnamefont{Salamon}},
  \bibinfo{author}{\bibfnamefont{T.}~\bibnamefont{Vertse}}, \bibnamefont{and}
  \bibinfo{author}{\bibfnamefont{L.}~\bibnamefont{Balkay}}
  (\bibinfo{year}{2012}), \eprint{1210.1721}.

\bibitem[{\citenamefont{Nagy et~al.}(2013)\citenamefont{Nagy, Fazekas, Juhasz,
  and Sailer}}]{Nagy:2013hka}
\bibinfo{author}{\bibfnamefont{S.}~\bibnamefont{Nagy}},
  \bibinfo{author}{\bibfnamefont{B.}~\bibnamefont{Fazekas}},
  \bibinfo{author}{\bibfnamefont{L.}~\bibnamefont{Juhasz}}, \bibnamefont{and}
  \bibinfo{author}{\bibfnamefont{K.}~\bibnamefont{Sailer}}
  (\bibinfo{year}{2013}), \eprint{1307.0765}.

\bibitem[{\citenamefont{Marian et~al.}(2014)\citenamefont{Marian, Jentschura,
  and Nandori}}]{Marian:2013zza}
\bibinfo{author}{\bibfnamefont{I.}~\bibnamefont{Marian}},
  \bibinfo{author}{\bibfnamefont{U.}~\bibnamefont{Jentschura}},
  \bibnamefont{and} \bibinfo{author}{\bibfnamefont{I.}~\bibnamefont{Nandori}},
  \bibinfo{journal}{J.Phys.} \textbf{\bibinfo{volume}{G41}},
  \bibinfo{pages}{055001} (\bibinfo{year}{2014}), \eprint{1311.7377}.

\bibitem[{\citenamefont{Reuter and Saueressig}(2002)}]{Reuter:2001ag}
\bibinfo{author}{\bibfnamefont{M.}~\bibnamefont{Reuter}} \bibnamefont{and}
  \bibinfo{author}{\bibfnamefont{F.}~\bibnamefont{Saueressig}},
  \bibinfo{journal}{Phys.Rev.} \textbf{\bibinfo{volume}{D65}},
  \bibinfo{pages}{065016} (\bibinfo{year}{2002}), \eprint{hep-th/0110054}.

\bibitem[{\citenamefont{Nandori}(2011)}]{Nandori:2010ij}
\bibinfo{author}{\bibfnamefont{I.}~\bibnamefont{Nandori}},
  \bibinfo{journal}{Phys.Rev.} \textbf{\bibinfo{volume}{D84}},
  \bibinfo{pages}{065024} (\bibinfo{year}{2011}), \eprint{1008.2934}.

\bibitem[{\citenamefont{Kov\'acs et~al.}(2013)\citenamefont{Kov\'acs, Nagy, and
  Sailer}}]{Kovacs:2013acta}
\bibinfo{author}{\bibfnamefont{J.}~\bibnamefont{Kov\'acs}},
  \bibinfo{author}{\bibfnamefont{S.}~\bibnamefont{Nagy}}, \bibnamefont{and}
  \bibinfo{author}{\bibfnamefont{K.}~\bibnamefont{Sailer}},
  \bibinfo{journal}{Acta Phys. Debr.} \textbf{\bibinfo{volume}{48}},
  \bibinfo{pages}{77} (\bibinfo{year}{2013}).

\bibitem[{\citenamefont{Nandori et~al.}(2014)\citenamefont{Nandori, Marian, and
  Bacso}}]{Nandori:2013nda}
\bibinfo{author}{\bibfnamefont{I.}~\bibnamefont{Nandori}},
  \bibinfo{author}{\bibfnamefont{I.}~\bibnamefont{Marian}}, \bibnamefont{and}
  \bibinfo{author}{\bibfnamefont{V.}~\bibnamefont{Bacso}},
  \bibinfo{journal}{Phys.Rev.} \textbf{\bibinfo{volume}{D89}},
  \bibinfo{pages}{047701} (\bibinfo{year}{2014}), \eprint{1303.4508}.

\end{thebibliography}

\end{document}